\documentclass[twocolumn,showpacs,preprintnumbers,amsmath,amssymb]{revtex4}
\usepackage{tipa}
\usepackage{amssymb}
\usepackage{stmaryrd}
\usepackage{txfonts}

\usepackage{graphicx}
\usepackage{dcolumn}
\usepackage{bm}

\begin{document}

\preprint{}

\title{Modified weak measurements for detecting photonic spin Hall effect}
\author{Shizhen Chen}
\author{Xinxing Zhou}
\author{Chengquan Mi}
\author{Hailu Luo}\email{hailuluo@hnu.edu.cn}
\author{Shuangchun Wen}
\affiliation{Laboratory for Spin Photonics, School of Physics and
Electronics, Hunan University, Changsha 410082, China}
\date{\today}

\begin{abstract}
Weak measurement is an important technique for detecting the tiny
spin-dependent splitting in photonic spin Hall effect. The weak
measurement is only valid when the probe wavefunction remains almost
undisturbed during the procedure of measurements. However, it does
not always satisfy such condition in some practical situations, such
as in the strong-coupling regime or the preselected and postselected
states are nearly orthogonal. In this paper, we develop a modified
weak measurement for detecting photonic spin Hall effect when the
probe wavefunction is distorted. We find that the measuring
procedure with preselected and postselected ensembles is still
effective. This scheme is important for us to detect the photonic
spin Hall effect in the case where neither weak nor strong
measurements can detect the spin-dependent splitting. The modified
theory is valid not only in weak-coupling regime but also in the
strong-coupling regime, and especially in the intermediate regime.
The theoretical models of conventional weak measurements and
modified weak measurements are established and compared. We show
that the experimental results coincide well with the predictions of
the modified theory.
\end{abstract}

\pacs{03.65.Ta, 42.25.-p, 42.25.Hz}
\keywords{photonic spin Hall effect, weak measurements, spin-orbit
coupling}

\maketitle

\section{Introduction}\label{SecI}
Weak measurements as an extension of quantum measurements were first
introduced by Aharonov, Albert, and Vaidman~\cite{Aharonov1988}. In
quantum measurements, the observable of system couples a probe state
with a pointer whose value can be read out by a meter. In general,
the conventional quantum measurements involve in a process of strong
coupling with the probe wavefunction is distorted. Weak measurements
suggest that the coupling between the observable and the probe state
is weak and the probe wavefunction remains almost undisturbed. The
weak value of an observable $\hat{A}$ outside the eigenvalue
spectrum can be obtained and the results are much larger than any
eigenvalues of the quantum system. It is shown that the weak value
$A_{w}$ can be formed as a simple expression
\begin{equation}
A_{w}=\frac{\langle\psi_{f}|\hat{A}|\psi_{i}\rangle}{\langle\psi_{f}|\psi_{i}\rangle}\label{asi},
\end{equation}
in which $|\psi_{i}\rangle$ and $|\psi_{f}\rangle$ are the
preselected and postselected states,
respectively~\cite{Aharonov1988,Duck1989,Ritchie1991,Parks1998}.
Indeed, the weak measurements have become a useful tool for
high-precision measurements of small physical parameters, such as
single-photon tunneling time~\cite{Steinberg1993}, deflections of
light beam~\cite{Dixon2009}, phase shift~\cite{Brunner2010},
frequency shift~\cite{Starling2010}, single-photon
nonlinearity~\cite{Feizpour2011}, high-resolution phase
estimation~\cite{Xu2013}, and angular
rotations~\cite{Magana-Loaiza2014}. In addition, it also assists us
in researching fundamental questions of quantum mechanics such as
single photon's polarization\cite{Pryde2005}, Hardy's
Paradox~\cite{Lundeen2009}, photon trajectories~\cite{Kocsis2011},
Heisenberg's uncertainty principle~\cite{Rozema2012}, quantum
polarization state~\cite{Salvail2013}, direct measurement of the
quantum wavefunction~\cite{Lundeen2011,Mirhosseini2014},
high-dimensional state vector~\cite{Malik2014}, and quantum Cheshire
cat~\cite{Aharonov2013,Denkmayr2014}.

As one of important applications, Hosten and Kwiat develop a weak
measurement to detect a tiny spin-dependent splitting in photonic
spin Hall effect (SHE)~\cite{Hosten2008}. Such an effect is
attributed to spin-orbit interaction and implied by angular momentum
conservation~\cite{Onoda2004,Bliokh2006}. In the procedure of weak
measurements, the quantum system is first preselected as a initial
state. Then the observable is very weekly coupled the pointer state.
Finally, the pointer position is recorded when the quantum system is
postselected in a final state. Weak measurements are valid only in
the regime of weak coupling between the observable and the probe
state~\cite{Krowne2009}. However, it does not always satisfy such
condition in some practical situations. The conventional weak
measurements should be modified if the coupling strength is not weak
enough~\cite{Wu2011,Zhu2011,Koike2011,Dressel2012I,Lorenzo2012,Pan2012}.
In addition, when the $|\psi_{i}\rangle$ and $|\psi_{f}\rangle$ are
nearly orthogonal, $\langle\psi_{f}|\psi_{i}\rangle\rightarrow0$,
the weak value $A_{w}$ can become arbitrary large. In fact, the weak
value $A_{w}$ should be modified in this
situation~\cite{Geszti2010,Nakamura2012,Pang2012,Kofman2012}. Note
that the probe wavefunction is distorted in these two cases, and
neither conventional weak nor strong measurements can detect the
spin-dependent splitting in photonic SHE.

In this paper, we develop a modified weak measurement for detecting
photonic SHE when the probe wavefunction is distorted. We consider
two possible cases leading to the distortion in the process of
measurements: one is due to the strong coupling, the other to the
preselected and postselected states are nearly orthogonal. We find
that the measuring procedure with preselected and postselected
ensembles is still effective when the probe wavefunction is strongly
distorted. The paper is organized as follows: In Sec. II, both the
conventional and the modified weak measurements for detecting the
photonic SHE are established. Subsequently, the evolution of the
wavefunction with different preselected and preselected states in
the weak measurements is analyzed in detail. In Sec. III, the
experimental and theoretical results are compared and discussed. In
contrast to the value of the conventional theory, our experimental
data agree well with the modified theory. In Sec. IV, a summary is
given.

\section{Theoretical model}\label{SecII}
In this section, we develop a modified theoretical model of the weak
measurements for detecting photonic SHE. As a comparison, the
conventional weak measurements are also reviewed. In quantum system
of the weak measurements, an initial state is first prepared. After
the system is weakly coupled with measuring device, the observable
$\hat{A}$ undergos a separate degree which is interpreted as a
meter. Then we read out the information from it when the
postselected state is performed. Here, the transverse spatial
distribution of light is used as a meter and the observable is
$\hat{\sigma}_3$. For simplicity, we only consider the preselected
states $|H\rangle$ and $|V\rangle$. With spin basis $|+\rangle$ and
$|-\rangle$, we have the expressions $|H\rangle=(|+\rangle +
|-\rangle)/{\sqrt{2}}$ and $|V\rangle=i(|-\rangle
-|+\rangle)/{\sqrt{2}}$. As an example, we consider the photonic SHE
in reflection at air-glass interface. To start, in our weak
measurements, the initial state $|H\rangle$ is first preselected. We
only consider the packet spatial extent in $y$ direction, and the
total wavefunction can be written as
\begin{align}
|\psiup_{initial}\rangle&=\int d y\psi(y)|y\rangle|\psi_{i}\rangle
=\int d k_{y} \phiup(k_{y})|k_{y}\rangle|\psi_{i}\rangle\nonumber\\
&=\int d k_{y} \phiup(k_{y})|k_{y}\rangle|H\rangle\label{asy},
\end{align}
where $\phiup(k_{y})$ is the Fourier transform of $\psi(y)$. We
assume that $\phiup(k_{y})$ is Gaussian spatial distribution here.
At the interface the light beam separates into two wave packets of
orthogonal spin states~\cite{Luo2011I}
\begin{align}
|k_{y}\rangle|H\rangle&\rightarrow|k_{y}\rangle(|H\rangle-k_{y}\delta^{H}|V\rangle)\nonumber\\
&=|k_{y}\rangle[\exp(+ik_{y}\delta^{H})|+\rangle+\exp(-ik_{y}\delta^{H})|-\rangle]/{\sqrt{2}}\label{asl},
\end{align}
where $\delta^{H}$ is given by
\begin{equation}
\delta^{H}=\frac{(r_{p}+r_{s})\cot\theta_{i}}{k_{0}r_{p}}\label{dh}.
\end{equation}
Here, $r_{p}$ and $r_{s}$ represent the Fresnel reflection
coefficients for parallel and perpendicular polarizations, respectively.
$\theta_{i}$ denotes the incident angle and $z_{r}$ is the Rayleigh length.

Taking the interaction Hamiltonian $\hat{H}=k_{y}\hat{A}\delta^{H}$
into account on reflection, the initial state becomes
\begin{equation}
|\psiup^{'}\rangle=\int d k_{y}
\phiup(k_{y})|k_{y}\rangle\exp(-ik_{y}\hat{A}\delta^{H})|\psi_i\rangle\label{asj}.
\end{equation}
In the weak measurements of photonic SHE, the meter states
corresponding to observable states $|+\rangle$ and $|-\rangle$
remain overlap. That is $|\delta^{H}|\ll w$, in which $w$ is the
width of the wavefunction. Under such condition, we expand the
operator $\exp(-ik_{y}\hat{A}\delta^{H})$ as
$1-ik_{y}\hat{A}\delta^{H}$. Therefore,
\begin{equation}
|\psiup^{'}\rangle\approx\int d k_{y}
\phiup(k_{y})|k_{y}\rangle|\psi_i\rangle-i \delta^{H}\int d k_{y}
\phiup(k_{y})k_{y}|k_{y}\rangle\hat{A}|\psi_i\rangle\label{asu}.
\end{equation}
With the relation of Eqs.~(\ref{asj}) and ~(\ref{asu}), the meter
state after postselection evolves as
\begin{align}
\langle\psi_f|\psiup^{'}\rangle&=\langle\psi_f| \left\{\int d k_{y}
\phiup(k_{y})|k_{y}\rangle\exp(-ik_{y}\hat{A}\delta^{H})|\psi_i\rangle\ \right\} \nonumber\\
&\approx\langle\psi_{f}|\psi_{i}\rangle\int d k_{y}
\phiup(k_{y})|k_{y}\rangle\left[1-ik_{y}\delta^{H}
\frac{\langle\psi_{f}|\hat{A}|\psi_{i}\rangle}{\langle\psi_{f}|\psi_{i}\rangle}\right]\nonumber\\
&=\langle\psi_{f}|\psi_{i}\rangle\int d k_{y}
\phiup(k_{y})|k_{y}\rangle[1-ik_{y}\delta^{H} A_w]\nonumber\\
&\approx\langle\psi_{f}|\psi_{i}\rangle\int d k_{y}
\phiup(k_{y})|k_{y}\rangle e^{-ik_{y}\delta^{H} A_w}\nonumber\\
&=\langle\psi_{f}|\psi_{i}\rangle\int d y \psi(y-\delta^{H}
A_w)|y\rangle\label{ass},
\end{align}
where
$A_w=\frac{\langle\psi_{f}|\hat{A}|\psi_{i}\rangle}{\langle\psi_{f}|\psi_{i}\rangle}$
is the conventional formalism of the weak value which is the same as
Eq.~(\ref{asi}). From the restrictions pointed out in
~\cite{Duck1989}, the validity of above calculation requires
\begin{equation}
|\delta^{H}A_w|\ll w\label{wvc}
\end{equation}
and
\begin{equation}
|\delta^{H}|/w\ll\min_{n=2,3,\ldots}\left|\frac{\langle\psi_{f}|\hat{A}|\psi_{i}\rangle}
{\langle\psi_{f}|\hat{A}^{n}|\psi_{i}\rangle}\right|^{1/(n-1)}\label{wvq}.
\end{equation}
The preselected state $|\psi_{i}\rangle$ here is the pure
polarization state $|H\rangle$ and the postselected state is
$|\psi_{f}\rangle=|V+\Delta\rangle$, in which $\Delta$ is referred
to as the postselected angle. That is
\begin{align}
&|\psi_{i}\rangle=|H\rangle,\\
&|\psi_{f}\rangle=\sin(-\Delta)|H\rangle+\cos(\Delta)|V\rangle\label{asm}.
\end{align}
In the spin basis, they become
\begin{align}
|\psi_{i}\rangle&=\frac{1}{{\sqrt{2}}}(|+\rangle + |-\rangle),\\
|\psi_{f}\rangle&=-\frac{i}{{\sqrt{2}}}(e^{- i \Delta}|+\rangle -
e^{+ i \Delta}|-\rangle)\label{asn}.
\end{align}
The operator $\hat{A}$ between these two states is $\hat{\sigma_3}$
since we deal with the left- and right-handed circularly
polarization basis. By calculating the matrix elements, we obtain
the weak value:
\begin{equation}
A_w=\frac{\langle\psi_{f}|\hat{\sigma_3}|\psi_{i}\rangle}{\langle\psi_{f}|\psi_{i}\rangle}=-
i\cot\Delta\label{asa},
\end{equation}
and from Eqs.~(\ref{wvc}) and ~(\ref{wvq}), the results is valid if
\begin{equation}
|\delta^{H}|/w\ll\min[\tan\Delta,\cot\Delta]\label{tlq}.
\end{equation}
In general, the weak value is a complex number of which the real and
imaginary parts correspond to the shifts of the position and
momentum in the wavefunction, respectively~\cite{Dressel2014}. It is
manifested experimentally in the plasmonic spin Hall
effect~\cite{Gorodetski2013}. Here, the pure imaginary weak value
$A_w$ converts the position displacements $\delta^{H}$ into a
momentum shift~\cite{Hosten2008,Dressel2012II}. The significance of
weak value is also attempted to be further understood in recent
works
~\cite{Aharonov2002,Jozsa2007,Williams2008,Romito2008,Kobayashi2012}.

We next consider the free evolution of the wavefunction before
detection. The displacement of the meter can be describe as
$A_w^{H_{con}}=F|A_w|$ in which the factor $F$ depends on the meter
state and free evolution~\cite{Aiello2008}. At any given plane $z$,
the free evolution factor is given by $F=z/z_{r}$, so we finally
obtain the amplified shift of the conventional theory as
\begin{align}
A_w^{H_{con}}\delta^{H}&=F|A_w|\delta^{H}\nonumber
\\&=\frac{z(r_{p}+r_{s})\cot\theta_{i}\cot\Delta}{z_{r}k_{0}r_{p}}\label{asz},
\end{align}
where $A_w^{H_{con}}$ is also defined as the conventional amplified
factor here. Equation~(\ref{asz}) suggests that in conventional
theory, the amplified shift, as well as the amplified factor
$A_w^{H_{con}}$, is proportional to the absolute value of the weak
value.

\begin{figure}
\includegraphics[width=8.5cm]{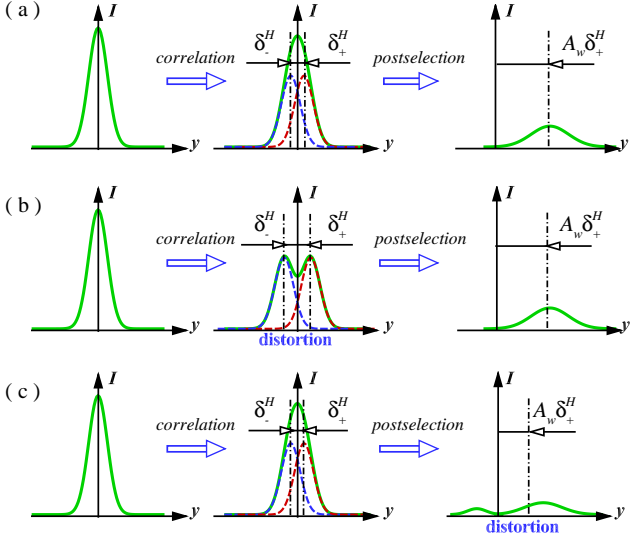}
\caption{\label{Fig1} (Color online) Schematic of quantum
measurements with preselected and postselected ensembles when probe
wavefunction is undisturbed (a) or distorted (b) and (c). System is
preselected an initial state $|\psi_{i}\rangle$ and its wavefunction
exhibits a Gaussian profile. After weak coupling with the system,
the wavefunction splits into two spin components(red dotted line and
blue dotted line, respectively). ${+}\delta$ and ${-}\delta$
represent the transverse shifts of left- and right-circularly
polarized components, respectively. Then we get the final pointer
position proportional to $A_{w}\delta$ after postselection. The
green solid lines indicate the total wave shapes. (a) The profile of
wavefunction is always Gaussian in the procedure of weak
measurement. The conventional weak measurements are valid in this
case. (b) The profile of probe wavefunction is distorted in strong
coupling, where the displacement exhibits a magnitude on the order
of the beam width $\delta\approx{w}$. (c) The other case of the
distortion of the probe wavefunction. It occurs when the preselected
and postselected states are nearly orthogonal. The conventional weak
measurements are invalid and the modified theory should be developed
in the latter two cases.}
\end{figure}

We now consider the preselected state of $|V\rangle$. In a similar
way we get the same weak value of the observable $A_w$. So we have
the amplified factor of the initial state $|V\rangle$ as
\begin{equation}
A_w^{V_{con}}=A_w^{H_{con}}=F|A_w|=\frac{z}{z_{r}}\cot\Delta\label{awvc}.
\end{equation}
In this case, the original transverse shift is~\cite{Qin2009}
\begin{equation}
\delta^{V}=\frac{(r_{p}+r_{s})\cot\theta_{i}}{k_{0}r_{s}}\label{as},
\end{equation}
and the final amplified shift after free propagation is obtained as
\begin{equation}
A_w^{V_{con}}\delta^{V}=\frac{z(r_{p}+r_{s})\cot\theta_{i}\cot\Delta}{z_{r}k_{0}r_{s}}\label{asp}.
\end{equation}

From the above analysis, it should be noted that the conditions to
obtain the weak value of conventional formalism are too strict. In
the experiment of the photonic SHE, if preselected and postselected
states are nearly orthogonal, indicating
$\langle\psi_{f}|\psi_{i}\rangle\rightarrow0$, the $A_w$ is very
large. Thus, the approximations in Eq.~(\ref{ass}) is invalid:
$[1-ik_{y}\delta^{H,V} A_w]\nthickapprox e^{-ik_{y}\delta^{H,V}
A_w}$. On the other hand, supposing that the preselected state
$|H\rangle$ is incident near the Brewster angle, the operator
$\exp(-ik_{y}\hat{A}\delta^{H})$ cannot be expanded as
$1-ik_{y}\hat{A}\delta^{H}$ due to the strong
coupling~\cite{Luo2011II,Kong2012,Pan2013,Gotte2013}. Therefore, the
weak value $A_w$ in Eq.~(\ref{ass}) is inaccurate if one of the
conditions is not satisfied. In fact, the two approximations above
require the restriction $\delta^{H,V} \cot\Delta\ll w$ from
Eq.~(\ref{tlq}).

We next calculate the final state of the meter in Eq.~(\ref{ass}) to
second order:
\begin{align}
|\phiup^{'}\rangle&=\langle\psi_f|\psiup^{'}\rangle
\nonumber\\&=\langle\psi_f| \left\{\int d k_{y}
\phiup(k_{y})|k_{y}\rangle\left(1-ik_{y}\delta\hat{A}-\frac{k_{y}^{2}\delta^{2}}{2}\hat{A}^{2}+\cdots\right)|\psi_i\rangle\ \right\} \nonumber\\
&\approx\langle\psi_{f}|\psi_{i}\rangle\int d k_{y}
\phiup(k_{y})|k_{y}\rangle\left(1-ik_{y}\delta
A_w-\frac{k_{y}^{2}\delta^{2}}{2}A_w^{2}\right)\label{aji},
\end{align}
where
$A_w^{2}=\frac{\langle\psi_{f}|\hat{A}^{2}|\psi_{i}\rangle}{\langle\psi_{f}|\psi_{i}\rangle}$
is the second-order weak value and $\delta$ is the transverse shift
$\delta^{H}$ or $\delta^{V}$. The expectation value of the position
is written as
\begin{align}
\langle y\rangle&=\frac{\langle\phiup^{'}|y|\phiup^{'}\rangle
}{\langle\phiup^{'}|\phiup^{'}\rangle }\nonumber
\\&=-\frac{2w^{2}\delta(2w^{2}+\delta^{2})\mathrm{Re}(A_w)}{4w^{2}+3\delta^{4}+4w^{2}\delta^{2}(|A_w|^{2}-1)},\label{fcg}
\end{align}
with the property $\hat{A}^{2}=1$. Here, the $A_w$ is imaginary and
therefore we consider the particular form of $\phiup(k_{y})$ with
the effective propagation distance $z$:
$\phiup(k_{y})\rightarrow\phiup_{z}(k_{y})=\phiup(k_{y})\mathrm{exp}\left(-i\frac{k_{y}^{2}}{2k_{0}}z\right)$~\cite{Hosten2008}.
Then we can recalculate the position of second-order theory as
\begin{equation}
\langle
y\rangle_{2nd}=\frac{4z\delta(3\delta^{2}-2w^{2})|A_w|}{k_{0}(4w^{4}+3\delta^{4}+4w^{2}\delta^{2}(|A_w|^{2}-1))}\label{sec}.
\end{equation}

In the following, we consider the modified theory without
approximation. The measuring procedure with preselected and
postselected ensembles is still effective in the modified weak
measurements. For the preselected state $|H\rangle$, the wave vector
$k_{x}$ should be reconsidered when the wave packet incident near
the Brewster angle. The evolution in the state after reflection can
be written as
\begin{align}
|k_{x,y}\rangle|H\rangle&\rightarrow|k_{x,y}\rangle|\varphi\rangle\nonumber\\ &=|k_{x,y}\rangle
\left[\left(r_{p}-\frac{k_{x}}{k_{0}}\chi \right)|H\rangle-k_{y}r_{p}\delta^{H}|V\rangle\right]\nonumber\\
&=|k_{x,y}\rangle\sum\limits_{+,-}\frac{1}{{\sqrt{2}}}
\left(r_{p}-\frac{k_{x}}{k_{0}}\chi \pm ik_{y}r_{p}\delta^{H}\right)|\pm\rangle\nonumber\\
&\rightarrow|y\rangle\sum\limits_{+,-}(y\mp\delta^{H_{mod}})|\pm\rangle\label{esar},
\end{align}
where $\delta^{H_{mod}}$ is defined as the original transverse shift
and is given by
\begin{equation}
\delta^{H_{mod}}=\frac{2r_{p}z_{r}(r_{p}+r_{s})\cot\theta_{i}}{2k_{0}r_{p}^{2}z_{r}+\chi^{2}}\label{abj},
\end{equation}
and $\chi=\partial r_{p}/\partial\theta_{i}$. So the meter state can
be described as
\begin{align}
|\psiup^{'}\rangle=\int d k_{x}d k_{y}
\phiup_{z}(k_{x,y})|k_{x,y}\rangle |\varphi\rangle\label{mstate}.
\end{align}
Subsequently, combine Eq.~(\ref{asm}) with Eq.~(\ref{mstate}), the
final meter state becomes
\begin{align}
|\phiup^{'}\rangle&=\langle\psi_{f}|\psiup^{'}\rangle\nonumber
\\&=\int d k_{x}d k_{y} \phiup_{z}(k_{x,y})|k_{x,y}\rangle
\left[\left(\frac{k_{x}}{k_{0}}\chi-r_{p}
\right)\sin{\Delta}-k_{y}r_{p}\delta^{H}\cos{\Delta}\right]\label{fms}.
\end{align}
The expectation value of the pointer observable $y$, also referred
to as modified amplified shift, is obtained as
\begin{align}
\langle
y^{H}\rangle&=A_w^{H_{mod}}\delta^{H_{mod}}=\frac{\langle\phiup^{'}|y|\phiup^{'}\rangle
}{\langle\phiup^{'}|\phiup^{'}\rangle }\nonumber
\\&=\frac{z[2k_{0}
r_{p}z_{r}(r_{p}+r_{s})+\chi^{2}]\sin(2\Delta)\cot\theta_{i}}{2k_{0}z_{r}(r_{p}+r_{s})^{2}\cos^{2}
\Delta\cot^{2}\theta_{i}+4k_{0}^{2}r_{p}^{2}z_{r}^{2}\sin^{2}\Delta}\label{asf}.
\end{align}
As the preselected state $|V\rangle$, because the coupling is always
weak, it still becomes
\begin{align}
|k_{y}\rangle|V\rangle&\rightarrow|k_{y}\rangle(|V\rangle+k_{y}\delta^{V}|H\rangle)\nonumber\\
&=|k_{y}\rangle[\exp(+ik_{y}\delta^{V})|+\rangle+\exp(-ik_{y}\delta^{V})|-\rangle]/{\sqrt{2}}\label{kyv}.
\end{align}
With the similar calculation in Eqs.~(\ref{mstate})-(\ref{asf}), we
get the $\langle y^{V}\rangle$ as
\begin{equation}
A_w^{V_{mod}}\delta^{V}=\frac{z
r_{s}(r_{p}+r_{s})\sin(2\Delta)\cot\theta_{i}}{(r_{p}+r_{s})^{2}\cos^{2}
\Delta\cot^{2}\theta_{i}+2k_{0}r_{s}^{2}z_{r}\sin^{2}\Delta}\label{asd}.
\end{equation}
Here, $A_w^{H_{mod}}$ and $A_w^{V_{mod}}$ are the modified amplified
factors of states $|H\rangle$ and $|V\rangle$, respectively.
Equations~(\ref{asf}) and ~(\ref{asd}) imply that the theoretical
output value is no longer proportional to the weak value, which is
different from the conventional theory. But it is worth remarking
that the results of the modified theory can reduce to the
conventional results as if the condition of weak coupling is
satisfied.

For the preselected state $|V\rangle$, when the postselected angle
$\Delta$ is not too small, the term
$(r_{p}+r_{s})^{2}\cos^{2}\Delta\cot^{2}\theta_{i}$ in
Eq.~(\ref{asd}) can be ignorable, and Eq.~(\ref{asd}) returns to
Eq.~(\ref{asp}). But for the case of $|H\rangle$, except for the
postselected angle limit, the neglect of the term
$2k_{0}z_{r}(r_{p}+r_{s})^{2}\cos^{2}\Delta\cot^{2}\theta_{i}$ in
Eq.~(\ref{asf}) requires that the incident angle is far from the
Brewster angle. As a result, Eq.~(\ref{asf}) can be reduced to
Eq.~(\ref{asz}) (under such condition, the $\chi$ can be ignorable).
Similarly, such analysis also holds for simplifying the amplified
factors of the two different theories. Additionally, we point out
that the modified weak measurements are also needed for detecting
the photonic SHE with an arbitrary linearly polarized state.

If the condition of weak coupling is satisfied, a weak measurement
of the photonic SHE performs as Fig.~\ref{Fig1}(a). The distribution
of the wavefunction is always the Gaussian shape during the
procedure of weak measurement. Now we consider the preselected state
is $|H\rangle$ with incident angle near the Brewster angle. The
separation between two spin components becomes large, which causes
the distortion of the wavefunction [ Fig.~\ref{Fig1}(b)]. On the
other hand, in Fig.~\ref{Fig1}(c), when the preselected and
postselected states are nearly or exactly orthogonal, the distortion
also occurs after postselection. The distortion of the prove
wavefunction accompanies the violation of the limited condition
because in the conventional theory where the weak value is
calculated under the assumption that wavefunction remains Gaussian.
Hence, if the distortion occurs, the quantum system is in the regime
where at least one of the two conditions above is violated, and the
conventional theory is invalid. To verify it, we measure the
intensity of the beam in our weak measurement for some special
cases. Our experimental results are also compared with the
conventional theory during the discussion. The modified theory is
valid not only in weak-coupling regime but also in the
strong-coupling regime, and especially in the intermediate regime.

\section{Experimental results and discussions}\label{SecIII}

To perform the weak measurements of the photonic SHE, the coordinate
frame and the experimental setup are similar to that in
Ref.~\cite{Luo2011II}. A incident Gauss beam is generated by a He-Ne
laser. The two nearly crossed polarizers are used to select the
initial state $|\psi_{i}\rangle$ and final state $|\psi_{f}\rangle$.
In the experiment, we chose the preselected state as $|H\rangle$ or
$|V\rangle$, and the corresponding postselected state is
$|V+\Delta\rangle$ or $|H+\Delta\rangle$. The two lenses are used to
focus and collimate the beam. When the light beam impinges on
air-glass interface, the tiny spin-dependent splitting takes place.
After the light passes through the second lens, the amplified shift
is detected by a charge-coupled device (CCD).

\begin{figure}
\includegraphics[width=8cm]{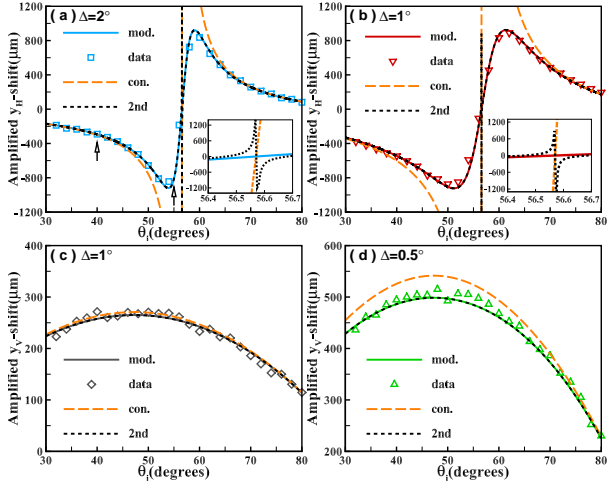}
\caption{\label{Fig2} (Color online) Amplified shifts vary with
incidence angles. (a) and (b) show the amplified shifts for
preselected state $|H\rangle$ with the fixed postselected angles
$\Delta$=$2^\circ$ and $1^\circ$, respectively. (c) and (d) are the
cases of $|V\rangle$ with $\Delta=1^\circ$ and $0.5^\circ$,
respectively. The solid lines are the modified values by
Eqs.~(\ref{asf}) and ~(\ref{asd}). In contrast, the long-dashed
lines are the predictions of conventional theory by Eqs.~(\ref{asz})
and ~(\ref{asp}), which have no upper bound for preselected state
$|H\rangle$ near the Brewster angle. And the short-dashed lines are
the result of second-order theory from Eq.~(\ref{sec}). Insets in
(a) and (b) show the difference of the three theories in
strong-coupling regime. The experimental data represented by hollow
points are also given. The arrows in the Fig.~\ref{Fig2}(a) indicate
the two cases in which we measure output intensity of the light
beams in Fig.~\ref{Fig4}.}
\end{figure}

We first consider the condition of weak coupling. With a fixed
preselected angle, the coupling strength varies with incident
angles. And then, with the invariable coupling strength, we analyze
the other case where the preselected and postselected states are
nearly orthogonal. That is, the incident angle is fixed and the
outcome is shown as a function of postselected angles. We now
experimentally measure the amplified shift, amplified factor, and
original displacement of the left-circularly polarized component. We
measure the amplified shifts in the case of incident angles varying
from $30^\circ$ to $80^\circ$, as shown in Fig.~\ref{Fig2}. In each
case, the experimental results are also given and agree well with
the theoretical curves of modified values. For comparison, the
conventional and second-order theories are shown as dashed lines.
For the preselected state $|H\rangle$ [Figs.~\ref{Fig2}(a) and
~\ref{Fig2}(b)], the conventional values are close to the modified
values under the condition that the incident angles are far from the
Brewster angle. But for the incident angles near the Brewster angle,
due to the enhancement of spin-orbit interaction, the splitting of
two spin components is nearly the same scale as the width of the
Gaussian beam. It means that the weak-coupling condition is violated
and the Gaussian profile of wavefunction is distorted. Remarkably,
in the strong-coupling regime, the second-order theory also exhibits
large distinction with the modified theory.

\begin{figure}
\includegraphics[width=8cm]{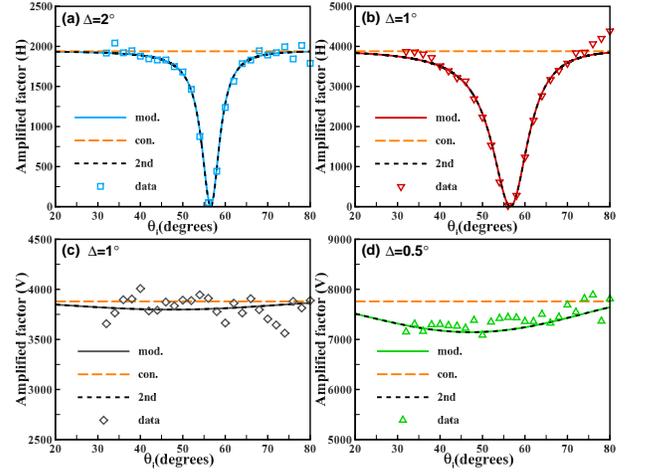}
\caption{\label{Fig3} (Color online) Amplified factors vary with
incidence angles. (a) and (b) show the cases of preselected state
$|H\rangle$. (c) and (d) are the preselected state $|V\rangle$. For
all panels, the long- and short-dashed curves are the conventional
and second-order theory from Eqs.~(\ref{awvc}) and Eqs.~(\ref{sec}),
respectively. The solid curves are modified values from
Eqs.~(\ref{asf}) and ~(\ref{asd}). The experimental data are
represented by hollow points.}
\end{figure}

It is interesting note that for different postselected angles, the
divergence between the conventional and modified theories is also
different. It indicates that the condition of postselected angle is
also important in the weak measurements as well as the weak-coupling
condition. In the case of preselected state $|V\rangle$ with the
certain angle $\Delta$, the amplifying shift varies with incident
angles as shown in Fig.~\ref{Fig2}(c), both the modified values and
the conventional values agree well with the experimental results.
Because there is not special angle like the Brewster angle for
preselected state $|V\rangle$, the interaction between the
observable and the probe state is so weak that the weak-coupling
condition is always satisfied. Therefore, as long as the angles
$\Delta$ are not too small, the conventional theory can be
approximatively equivalent to the modified theory. Otherwise, the
two theories diverge and there is almost no overlap between the
modified and conventional curves as shown in Fig.~\ref{Fig2}(d).

\begin{figure}
\includegraphics[width=8cm]{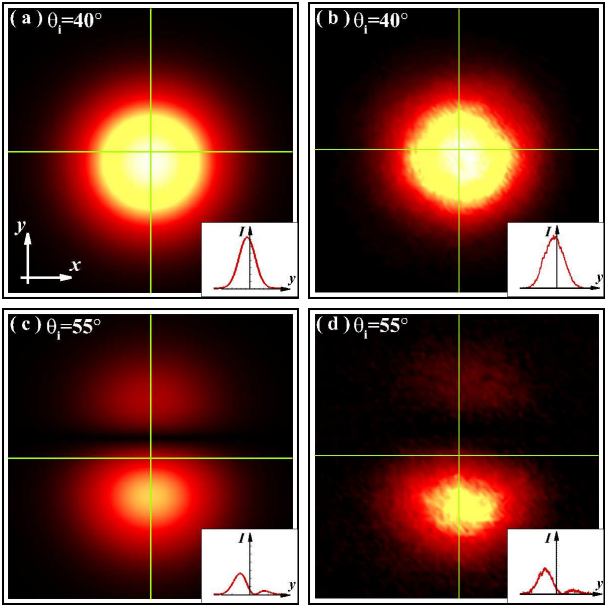}
\caption{\label{Fig4} (Color online) Distortion of wavefunction due
to the strong coupling. The initial state is preselected as
$|H\rangle$, the incident angles as $\theta_i=40^\circ$, and
$\theta_i=55^\circ$, respectively. In addition, the postselected
angles are both chosen as $\Delta=2^\circ$. (a) and (c) are the
theoretical prediction. (b) and (d) are the experimental results.
The insets indicate the intensity profiles along the $y$-axis. Note
that, in the regime of strong coupling $\theta_i=55^\circ$, the
Gaussian profile of wavefunction distorts and exhibits asymmetric
double-peak. But the wavefunction still remains Gaussian form with
$\theta_i= 40^\circ$.}
\end{figure}

To discuss the problem in detail, the corresponding amplified
factors shown in Fig.~\ref{Fig3} are also given. Generally,
according to Eq.~(\ref{awvc}), the amplified factor of the weak
measurements is a constant if the postselected angle $\Delta$ is
decided (the straight dashed lines). It is valid if two limited
conditions are all satisfied. But the modified amplified factors
$A_w^{H_{mod}}$ and $A_w^{V_{mod}}$ from Eqs.~(\ref{asf}) and
~(\ref{asd}) are not constants which are shown as solid curves,
which are identical to the curves of second-order theory. For state
$|H\rangle$, the amplified factor of the modified theory is
consistent with that in conventional theory when the incident angles
are far away from the Brewster angle, but it becomes small near the
Brewster angle. As a result, the largest divergence between modified
and conventional values occurs. Comparing Fig.~\ref{Fig3}(a) with
Fig.~\ref{Fig3}(b), we find that the postselected angle has also
important impact on the deviation of the two theories. As predicted,
the modified cures coincide well with our experimental data in the
strong-coupling regime. For the preselected state $|V\rangle$, we
only concern on the limit of postselected angle. In
Fig.~\ref{Fig3}(c), the conventional amplified factor approach to
the modified one when the angle $\Delta$ is chosen as $1^\circ$. But
at first glance, one may regard that there exists discrepancy
between the theory and measured data. As a matter of fact, such
deviation is appropriate because in this panel the spacing of
$y$-axis is suitably small in order to present the divergence of the
two theories. Figure~\ref{Fig3}(d) shows the distinction between the
two theories with the $\Delta=0.5^\circ$ which is small enough,
while our experimental results guarantee the validity of the
modified theory.

\begin{figure}
\includegraphics[width=8cm]{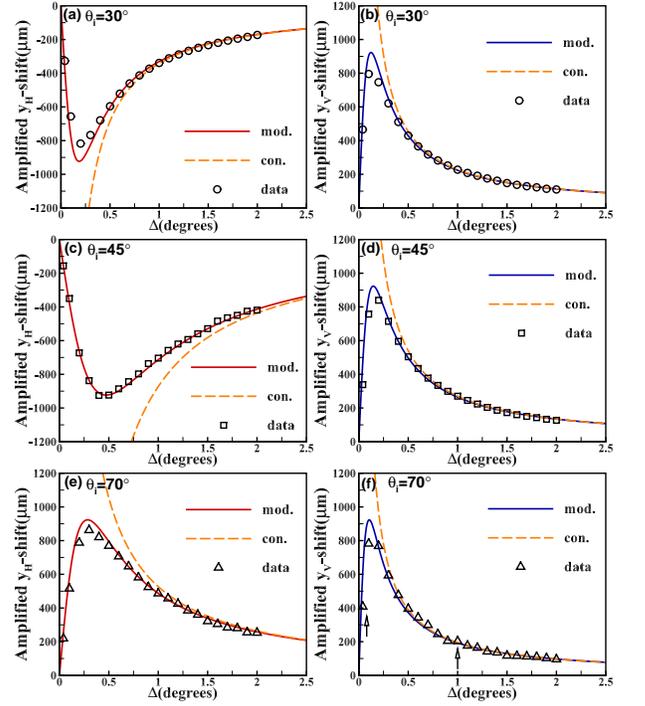}
\caption{\label{Fig5} (Color online) With the fixed incident angles
$\theta_i = 30^\circ$, $45^\circ$, and $70^\circ$, amplified pointer
shifts are shown as functions of postselected angle $\Delta$ for
preselected states $|H\rangle$ (left column) and $|V\rangle$ (right
column). Here, the modified theoretical values are represented by
solid lines from Eqs.~(\ref{asf}) and ~(\ref{asd}), and the values
in conventional theory are dashed lines from Eqs.~(\ref{asz}) and
~(\ref{asp}). The sharp value of the dashed lines is not shown in
each panel due to its infiniteness. The experimental data are
obtained as hollow points. The arrows in the Fig.~\ref{Fig5}(f) are
two special cases in which we measure their output intensity shown
in Fig.~\ref{Fig7}.}
\end{figure}

If the profile of Gaussian wavefunction is distorted, the theory of
weak measurements should be modified.  There are two possible cases
leading to the distortion in the process of measurements: one is due
to the strong coupling, the other to the preselected and
postselected states are nearly orthogonal. We first consider the
former case. The initial state is preselected as $|H\rangle$ as
shown in Fig.~\ref{Fig4}, which explains the connection between
distortion of the beam and the weak-interaction condition. To avoid
the limited condition of postselected angle, we set it to be
$\Delta=2^\circ$. We consider two special examples labeled by arrows
in Fig.~\ref{Fig2}(a). The measured intensity is read out from CCD
(right column of Fig.~\ref{Fig4}). For comparison, the corresponding
predictions are also given (left column of Fig.~\ref{Fig4}). The
case with incident angle $\theta_i=40^\circ$ suggests that the
conventional theory is equivalent to the modified one if the output
beam remains the Gaussian profile. Conversely, at the incident angle
$55^\circ$ the wavefunction distorts due to the strong coupling and
the conventional theory of weak measurements is invalid.

\begin{figure}
\includegraphics[width=8cm]{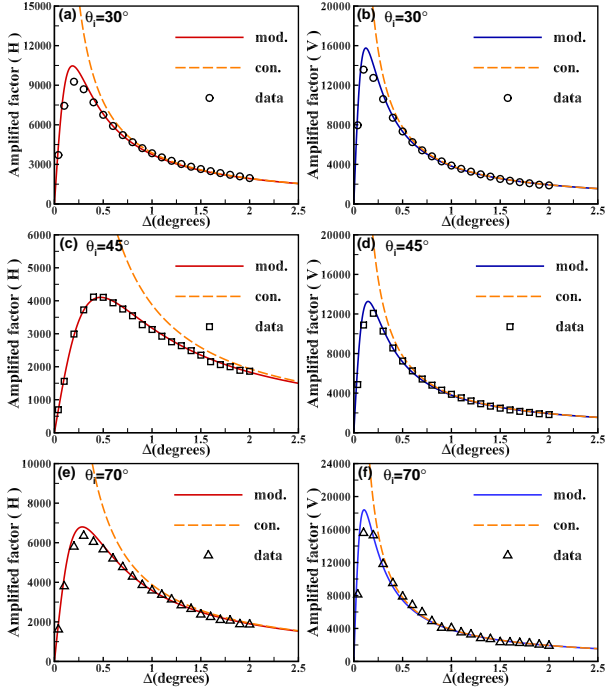}
\caption{\label{Fig6} (Color online) Amplified factors with various
postselected angle. (a), (c), and (e) show the cases of $|H\rangle$.
(b), (d), and (f) are $|V\rangle$. In term of Eqs.~(\ref{awvc}),
~(\ref{asf}), and ~(\ref{asd}), the predictions of the modified and
conventional theory are represented by solid lines and dashed lines,
respectively. The conventional amplified factor can be arbitrarily
large with the angle $\Delta$ decreasing to $0^\circ$. The hollow
points are experimental data.}
\end{figure}

All discussion above is about the weak-interaction condition in the
modified weak measurements. In the rest of the paper, we consider
another limited condition of small postselected angle, that is to
say the preselected and postselected states are nearly orthogonal.
We find that the weak measurements of conventional theory are also
not valid when the postselected angle $\Delta$ is too small. To
prove that, we detect the photonic SHE with the fixed incident
angles but various postselected angles. In the same sequence, we
first measure the amplified shifts with the angle $\Delta$ varying
from $0^\circ$ to $2.5^\circ$, which are shown in Fig.~\ref{Fig5}.
Under such condition, large divergence takes place between modified
and conventional theories when the angle $\Delta$ is close to
$0^\circ$. But the amplified values in conventional theory
essentially have no difference to that in modified theory with
appropriate postselected angles. In the conventional theory of the
weak measurements, according to the Eqs.~(\ref{asz}) and
~(\ref{asp}), the amplified value can be arbitrarily large if the
preselected and postselected states are nearly crossed, and such
value is shown as the dashed lines in the figures.

\begin{figure}
\includegraphics[width=8cm]{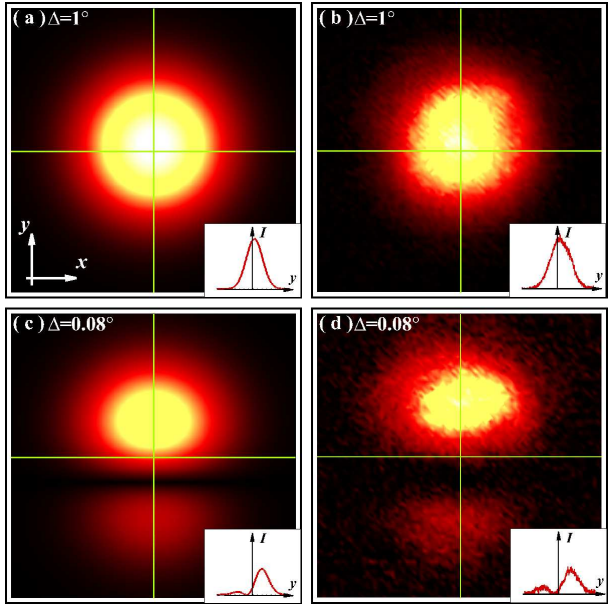}
\caption{\label{Fig7} (Color online) Distortion of probe
wavefunction due to the fact that the preselected and postselected
states are nearly orthogonal. We chose the preselected state as
$|V\rangle$ with a fixed incident angle at $\theta_i=70^\circ$. (a)
and (c) are the theoretical results. (b) and (d) are the
experimental results. The insets: the profiles of wavefunction at
the plane of $x=0$. It implies that the output wavefunction no
longer remains a Gaussian profile when the preselected and
postselected states are nearly orthogonal.}
\end{figure}

Whereas in our modified theory (the solid lines), with the angle
$\Delta$ continuously decreasing to $0^\circ$, the amplified shift
first increases and reaches the maximum value, then decreases
rapidly even to $0^\circ$. We do the experiment with the incident
angles $\theta_i = 30^\circ$, $45^\circ$, and $70^\circ$ both for
states $|H\rangle$ and $|V\rangle$, and our experimental data reveal
the trend like the modified theory. Note that in Fig.~\ref{Fig5}(c),
the modified and conventional curves begin to separate when the
$\Delta$ is near $2.5^\circ$, but in other cases, such separation
takes place until the $\Delta$ becomes smaller. The point is that in
this case the weak-interaction is a little strong at the incident
angle $\theta_i=45^\circ$. To clarify it, we see that the example in
Fig.~\ref{Fig5}(d), with the same incident angle but different
preselected state $|V\rangle$, dose not exist such problem. We point
out that when the postselected angles are negative, the values of
amplified shift are the same magnitude but opposite sign, besides,
the amplified shift of second-order theory is completely identical
to the modified one, which both are not shown here.

\begin{figure}
\includegraphics[width=8cm]{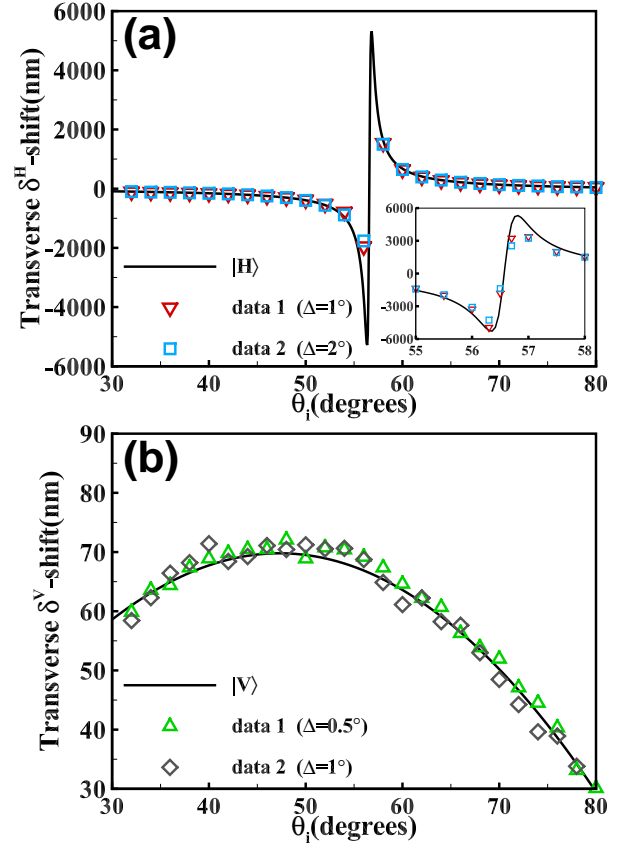}
\caption{\label{Fig8} (Color online) Original transverse shifts of
the spin component $|+\rangle$ for preselected states: (a)
$|H\rangle$ and (b) $|V\rangle$. The theoretical values are
represented by black solid lines from Eqs.~(\ref{as}) and
~(\ref{abj}). For each preselected state, the experiment data with
two different postselected angles $\Delta$ are shown as hollow
points. Inset in (a) shows the theoretical and experimental results
in strong-coupling regime.}
\end{figure}

In addition, the corresponding amplified factors are also obtained
in Fig.~\ref{Fig6}. We get the conventional amplified factors
represented by dashed lines in terms of the relation
$A_w^{H_{con}}=A_w^{V_{con}}=F|A_w|$. It is shown that the amplified
factor has no upper bound when the $\Delta$ approaches to $0^\circ$.
In fact, both for two preselected states $|H\rangle$ and
$|V\rangle$, the behavior of the amplified factors $A_w^{H_{mod}}$
and $A_w^{V_{mod}}$ shown as the solid lines is similar to that of
amplified shifts. In each panel, the modified theory is compared
with the conventional theory, and once again, the difference between
conventional and modified values would be significant if the
preselected angle $\Delta$ is small. The deviation of the amplified
factors between the two theories is also determined by incident
angle, especially for $|H\rangle$. The experimental results agree
well with the modified values. There exists a peak value with an
optimal postselected angle, and the amplification effect disappears
when the preselected and postselected states are completely
orthogonal. That is to say, by adjusting the postselected angle
$\Delta$, one can obtain the maximum amplified factor and improve
the precision for measuring the photonic
SHE~\cite{Zhou2014I,Zhou2014II}. Note that the signal amplification
from weak measurements has been extensively studied recently, such
as optimal probe wavefunction of weak-value
amplification~\cite{Susa2012}, technical advantages for weak-value
amplification~\cite{Jordan2014}, and maximizing the output by weak
values and weak coupling~\cite{Lorenzo2014}. Note that there are
still some open questions about whether the weak value ampliation
can suppress technical noise~\cite{Knee2014,Ferrie2014}.

We have discussed the connection between the condition of
weak-interaction and the output intensity of the light in
Fig.~\ref{Fig4}. Here, we analyze the distortion associated with the
postselected angle. In the case of Fig.~\ref{Fig5}(f), we measured
its intensity with a decreasing angle $\Delta$, and found that the
intensity changes gradually from a single Gaussian into asymmetrical
double-peak intensity. Then the double-peak intensity becomes
symmetric when postselected angle is equal to $0^\circ$. In
particular, in Fig.~\ref{Fig7}, we present the intensity with two
postselected angles which are labeled by the arrows in
Fig.~\ref{Fig5}(f). Both for the predicted and measured intensity,
they remain the Gaussian form with the $\Delta=1^\circ$, but distort
with the $\Delta=0.08^\circ$. We draw a conclusion that with
different postselected angles, the experimental data do not fit to
the conventional theory when the profile of wavefunction distorts.
Therefore, one may judge weather the modified theory is equivalent
to the conventional one by observing the output intensity of the
light.

The tiny original shifts of the component of $|+\rangle$ obtained by
the modified theory for preselected states $|H\rangle$ and
$|V\rangle$ are presented in Fig.~\ref{Fig8}. For each preselected
states, we detect it with two different postselected angles. We find
that the original shift of $|H\rangle$ is indeed very large near the
Brewster angle: nearly the same scale as the width of probe state
$|{\delta^H}|\approx{w}$, As the result, the profile is strongly
distorted and the weak measurements approximations fail. For the
preselected state $|V\rangle$, the original shifts for various
incident angles are always much less than the width of pointer state
$|{\delta^V}|\ll{w}$. Both experimental results agree well with the
theoretical predictions.

It should be noted in the strong-coupling case the spin-dependent
splitting is very sensitive to the variation of physical parameters
and therefore has important applications in precision metrology,
such as measuring thickness of metal film~\cite{Zhou2012I},
identifying graphene layers~\cite{Zhou2012II}, determining the
strength of axion coupling in topological
insulators~\cite{Zhou2013}, and detecting of magneto-optical
constant of magneto-optical media~\cite{Qiu2014}. However, in this
regime both conventional weak measurements theory and its
second-order corrections cannot obtain the exact meter shifts as the
analysis above. Hence the modified weak measurement is important in
precision metrology.

Finally, it should be mentioned that photonic SHE manifests as
spin-dependent splitting of light, which corresponds to two types of
geometric phases: the Rytov-Vladimirskii-Berry phase associated with
the evolution of the propagation direction of light and the
Pancharatnam-Berry phase related to the manipulation with the
polarization state of light~\cite{Bliokh2008I}. In general, the
spin-dependent splitting due to the Rytov-Vladimirskii-Berry phase
is limited by a fraction of the wavelength, and can only be detected
by weak measurements~\cite{Hosten2008}. However, the spin-dependent
splitting due to the Pancharatnam-Berry phase can be large enough
for direct detection (strong measurements) without using the weak
measurement
technology~\cite{Bliokh2008II,Gorodetski2008,Shitrit2011,Shitrit2013,Ling2014,Ling2015}.
In addition, the rapidly varying phase discontinuities along a
metasurface, breaking the axial symmetry of the system, enable the
direct observation of the spin-dependent splitting~\cite{Yin2013}.
In the intermediate regime, the modified theory is important where
neither weak nor strong measurements can detect the spin-dependent
splitting.

\section{Conclusions}
In conclusion, we have developed a modified weak measurements for
the detection of the photonic SHE. Compared with the conventional
weak measurements, the amplified shift, amplified factor, and
original displacement for different preselected and postselected
states have been examined. The conventional theory for preselected
state $|H\rangle$ is invalid in the strong-coupling regime or when
the preselected and postselected states are nearly orthogonal. But
for preselected state $|V\rangle$ it is only limited by the latter
since the condition of weak coupling is always satisfied. We have
shown that the weak measurements for detecting photonic SHE need to
be modified when one of the condition is violated. This is due to
the fact that probe wavefunction is distorted in the case of strong
coupling or preselected and postselected states are nearly
orthogonal. Otherwise, the modified theory can reduce to the
conventional one beyond the two restrictions. We have found that the
measuring procedure with preselected and postselected ensembles is
still effective. This scheme is important for us to detect the
photonic SHE in the case where neither weak nor strong measurements
can detect the spin-dependent splitting. Our modified theory is
valid not only in weak-coupling regime but also in the
strong-coupling regime, and especially in the intermediate regime.
We believe that such problem may also exist in the weak measurements
of other quantum systems and would have possible applications in
precision measurements.

\begin{acknowledgements}
This research was supported by the National Natural Science
Foundation of China (Grants Nos. 11274106, 11474089).
\end{acknowledgements}

\end{document}